\documentclass[12pt, a4paper]{article}
\usepackage{graphicx}
\usepackage{latexsym}
\usepackage{amsmath}
\usepackage{amssymb}
\usepackage{epsfig}
\begin{document}
%\begin{titlepage}
\title{Proof of a decomposition theorem for symmetric tensors on
spaces with constant curvature}
\author{Norbert Straumann \\
Institute for Theoretical Physics University of Zurich,\\
Winterthurerstrasse 190, CH--8057 Zurich, Switzerland}
%\date{}
\maketitle
\begin{abstract}
In cosmological perturbation theory a first major step consists in the decomposition of the various perturbation amplitudes into scalar, vector and tensor perturbations, which mutually decouple. In performing this decomposition one uses -- beside the Hodge decomposition for one-forms -- an analogous decomposition of symmetric tensor fields of second rank on  Riemannian manifolds with constant curvature. While the uniqueness of such a decomposition follows from Gauss' theorem, a rigorous existence proof is not obvious. In this note we establish this for smooth tensor fields, by making use of some important results for linear elliptic differential equations.
\end{abstract}
%\vspace{0.5cm}
%Key words: Cosmological perturbations, elliptic equations, harmonic \\analysis.\\  PACS numbers: 98.80.Jk, 02.40.-k

%\end{titlepage}

\section{The decomposition theorem}\label{s1}

In cosmological perturbation theory one can regard the various
perturbation amplitudes as time dependent tensor fields on a
three-dimensional Riemannian space $(M,g)$ of constant curvature
$K$ (see, e.g., \cite{NS}). For skew-symmetric tensor fields (p-forms) there is on
arbitrary compact Riemannian manifolds the profound Hodge
decomposition into an orthogonal direct sum of exact, coexact, and
harmonic forms. No analogous decomposition for symmetric tensor
fields, say, is available in general. However, when the space has
constant curvature, a symmetric tensor field $t_{ij}$ can be
decomposed as follows:
\begin{equation}
t_{ij}=t^{(S)}_{ij} + t^{(V)}_{ij} + t^{(T)}_{ij} ~, \label{eq:1}
\end{equation}
where
\begin{eqnarray}
t_{ij}^{(S)} &= & \frac{1}{3}t^k{}_kg_{ij} + (\nabla_i\nabla_j -
\frac{1}{3}g_{ij}\nabla^2)f~,\label{eq:2}\\
t_{ij}^{(V)} &= & \nabla_i\xi_j + \nabla_j\xi_i,\label{eq:3}\\
t_{ij}^{(T)} &: &  t^{(T)i}{}_i=0;~~ \nabla_j
t^{(T)ij}=0.\label{eq:4}
\end{eqnarray}
In these equations $f$ is a function on $M$ and $\xi^i$ a vector
field  with vanishing divergence; $\nabla^2$  denotes
$g^{ij}\nabla_i\nabla_j$ on $(M,g)$. (Note that this does not agree
with the Laplace-Beltrami operator $\triangle$ for differential
forms, except on functions. The difference is given by the Weitzenb\"{o}ck
formula \cite{JJ}. For general tensor fields $\nabla^2$ is the
natural extension of the Laplace operator on functions.) The three
components are easily shown to be orthogonal to each other with
respect to the scalar product.
\begin{equation}
\langle t,s\rangle=\int_\Sigma t_{ij}s^{ij} d\mu ~, \label{eq:5}
\end{equation}
where $\mu$ is the Riemannian measure for the metric $g$. This fact
implies that the decomposition of $t_{ij}$ is unique. Below we give
a rigorous existence proof.

\subsection{Some tools}\label{s1.1}

In this subsection $(M,g)$ can be an arbitrary compact (closed)
Riemannian manifold. On this we consider operators
\begin{equation}
L=-\triangle + k, ~~ k\in \mathbb{R}. \label{eq:6}
\end{equation}
Specializing existence and regularity results from the theory of
elliptic partial differentials equations, established for instance
in chapter 5 of \cite{Tay}, the following holds:

(i) The equation $Lu=f$, with $f\in C^\infty(M)$ has a solution
$u\in C^\infty(M)$ if and only if $f$ is orthogonal to the smooth
functions $v$ satisfying $Lv=0$. In particular,
$\triangle(C^{\infty}(M))=H^{\bot}$: the orthogonal complement of
the harmonic functions $H$ in $C^{\infty}(M)$.

(ii) In the space of smooth functions the equation $Lu=f$ has always
a unique solution, if $k$ is not an eigenvalue of the operator
$\triangle$.

(iii) If $k$ is an eigenvalue of $\triangle$, and $f$ is orthogonal
to the smooth eigenfunctions $w$ of $\triangle$ with eigenvalue $k$,
then there are smooth solutions of $Lu=f$. Any two of them differ by
such an eigenfunction $w$.

In passing we note that \emph{$L^2$-completeness}, as well as
\emph{uniform completeness} of the smooth eigenfunctions of
$\triangle$ holds. We will, however, not use this fact. We also
recall that harmonic functions on $M$ are constant.

\subsection{Proof of the decomposition theorem}

Let now $(M,g)$ be a Riemannian space of constant curvature $K$.
Then the Ricci tensor and Ricci scalar are given by
\begin{equation}
R_{ij}=(n-1)Kg_{ij}, ~~ R=n(n-1)K. \label{eq:7}
\end{equation}
Below we shall use the following consequence of the Ricci identity:
\begin{equation}
\nabla^2\nabla_i\omega_j=\nabla_i\nabla^2\omega_j+
K[(n-1)\nabla_i\omega_j+2\nabla_j\omega_i-2g_{ij}\nabla^k\omega_k].\label{eq:8}
\end{equation}

For definiteness we consider the compact case of an n-dimensional
sphere. (In the non-compact case one has to add fall-off
conditions.)

The decomposition theorem follows immediately, once we have shown
that for any symmetric traceless tensor $t_{ij}$ there exists a
covariant vector field $A_i$, such that
\begin{equation}
t_{ij}-\nabla_iA_j-\nabla_jA+\frac{2}{n}\; g_{ij}\nabla^kA_k
\label{eq:9}
\end{equation}
is transversal, i.e., satisfies the second equation in (\ref{eq:4}).
(Apply in a second step the decomposition (\ref{eq:12}) below.)
With the help of the Ricci identity and (\ref{eq:7}) this condition
can be written as
\begin{equation}
[\nabla^2 +(n-1)K]A_i
+\left(1-\frac{2}{n}\right)\nabla_i(\nabla_jA^j)=\nabla^j
t_{ij}\;.\label{eq:10}
\end{equation}
So, the existence of a decomposition (\ref{eq:1}) is equivalent to
the question of whether there is a covariant vector field satisfying
equation (\ref{eq:10}). We now show that this question has a
positive answer.

Applying $\nabla^i$ on (\ref{eq:10}), and using as a special case
of (\ref{eq:8}) the identity $\nabla^i\nabla^2 A_i=\nabla^2
\nabla^iA_i +(n-1)K\nabla_iA^i$, we obtain
\begin{equation}
(\triangle+nK)\nabla^iA_i=\frac{n}{2(n-1)}\nabla^i\nabla^jt_{ij}\;.
\label{eq:11}
\end{equation}
As a special case of the Hodge decomposition, $A_i$ can be uniquely
decomposed into an direct orthogonal sum of the form
\begin{equation}
A_i=V_i+\nabla_iS, ~~ \nabla^iV_i=0, \label{eq:12}
\end{equation}
whence
\begin{equation}
\nabla^iA_i=\triangle S. \label{eq:13}
\end{equation}
Then (\ref{eq:11}) becomes
\begin{equation}
\triangle[(\triangle+nK)S]=\frac{n}{2(n-1)}\nabla^i\nabla^jt_{ij}
\;. \label{eq:14}
\end{equation}
Note that $\lambda_1:=-nK$ is an eigenvalue of $\triangle$. Since
the right-hand side of this equation is by Gauss' theorem in
$H^{\bot}$, equation (\ref{eq:14}) has, up to an additive constant,
a  unique solutions for $(\triangle+nK)S$. Equation (\ref{eq:10})
can be rewritten as
\begin{equation}
[\nabla^2 +(n-1)K]V_i =\nabla^j
t_{ij}-\frac{2(n-1)}{n}\nabla_i(\triangle S+nKS)\;. \label{eq:15}
\end{equation}

There are certainly solutions of (\ref{eq:14}) and (\ref{eq:15}).
For the latter one has to use property (ii) of Sect. 1.1 for
1-forms. The left-hand side of (\ref{eq:15}) is equal to the
operator $\triangle + 2(n-1)K$ applied on the 1-form belonging to
$V_i$. For any solution of the two equations, $A_i$ given by
(\ref{eq:12}) then satisfies equation (\ref{eq:10}). Indeed,
applying $\nabla^i$ on (\ref{eq:15}) and using (\ref{eq:14}) leads
to $[\triangle+2(n-1)K]\nabla^iV_i=0$, hence $\nabla^iV_i=0$. Then,
the definition (\ref{eq:12}) implies $\nabla^iA_i=\triangle S$. If
one now replaces $V_i$ in (\ref{eq:15}) by $V_i=A_i-\nabla_iS$ and
sets $\triangle S=\nabla^iA_i$ in the resulting equation, one
recovers (\ref{eq:10}).

This concludes the proof.

%%%%%%%%%%%%%%%%%%%%%%%%%%%%%%%%%%%%%%%%%%%%%%%%%%%%%%%%%%%%%%%%%%%%%%%%

%%%%%%%%%%%%%%%%%%%%%%%%%%%%%%%%%%%%%%%%%%%%%%%%%%%%%%%%%%%%%%%%%%%%%
\end{document}